\documentclass[a4paper,twocolumn,fleqn]{article}

 \usepackage{amsmath}             
 \usepackage[psamsfonts]{amssymb} 
\usepackage{helvet}
\usepackage{subfigure}

\usepackage[dvipdfm]{graphicx}     
\usepackage{jpfr}                  

\begin{document}

\title{Shear flow effects on double tearing mode global magnetic reconnection}


\author{Thibaut~VOSLION\sup{1,2,3}, Olivier~AGULLO\sup{1,2}, Peter~BEYER\sup{1,2}, Masatoshi~YAGI\sup{3}, Sadruddin~BENKADDA\sup{1,2}, Xavier~GARBET\sup{4}, Kimitaka~ITOH\sup{5}, Sanae-I.~ITOH\sup{3}}

\affiliation{
  \sup{1}France Japan Magnetic fusion Laboratory, LIA 336, france-Japan \\
  \sup{2}LPIIM, CNRS, Université de Provence, Marseille 13397, France\\
  \sup{3}RIAM,Kyushu University, Kasuga 6-1, Kasuga 816-8580, Japan\\
  \sup{4}CEA, IRFM, f-13108 Saint Paul Lez Durance, France\\
  \sup{5}National Institute for Fusion Science , Oroshi - cho, Toki -shi , 509-52 Japan}

\date{\ \ }


\begin{abstract}
The dynamics of a global reconnection in the presence of a poloidal shear flow which is located in between magnetic islands is investigated. Different linear regimes are identified according to the value of the resistivity and the distance between the low-order resonant surfaces. It is found that the presence of a small shear flow affects and significantly delays the global reconnection processes. It is shown that this delay is linked to a breaking of symmetry imposed by the existence of the shear flow and the generation of a mean poloidal flow in the resistive layers. 

\end{abstract}


\maketitle  

\section{\label{sec:1}Introduction}
In tokamaks, the internal transport barrier is observed in reversed magnetic shear configurations where two low-order rational surfaces exist \cite{Budny02}. 
In such configurations, toroidal and poloidal flows, as well as temperature and density gradients can coexist close to the plasma core  \cite{Wolf03}. 
Indeed, there is evidence that such zonal flows exist and that the shear flow is localized in between resonant surfaces where the double tearing instability can grow. The observed maxima of the flow velocity are weak compared with the Alfv\'en velocity $v_A$. Nevertheless,  such flows can be important with respect to the turbulence level because  the associated radial electric field can be linked to the formation of an ITB\cite{Bell98}. 
In fact, such configurations, 
allowing eventually the generation of a strong internal barrier, are part of the ITER scenarios for advanced confinement\cite{Sips04}. 

 Previous works have investigated the influence of a Bickley jet on a tearing instability\cite{Ofman91,Biskamp98} 
and on a double tearing instability\cite{Ofman92}. More recently, Bierwage {\it et al} \cite{Bierwage07} have studied the influence of the core rotation amplitude - for a given sheared flow profile - on the stability of magnetohydrodynamic (MHD) and Kelvin-Helmholtz instabilities (KHI). These studies are mainly linear and a cylindrical geometry is used.
They have shown that KHI can develop at large poloidal mode numbers and are strongly enhanced when the core rotation passes a critical value of the order of $10^{-2}v_A$ in the case where the distance between the resonant surfaces is small. They also found that below this critical value the growth rates  of both the MHD and KH mode decrease with the amplitude of the core velocity.\\
In this paper, we focus on the case where the shear flow is in  between the two tearing instabilities. Within the framework of MHD, a  2D slab geometry is used. It is found that the nature of the dynamics depends drastically on the distance between the low-order resonant surfaces. The impact of the flow on the nonlinear process leading to a global magnetic reconnection of the system is studied. The origin and the role of the dynamically generated $m=0$ poloidal flow turbulence are investigated as well as the dynamics at small scales once the reconnection has taken place.

\section{\label{sec:2}Model equations}
%
%
\begin{figure}
  \includegraphics[width=7cm,clip]{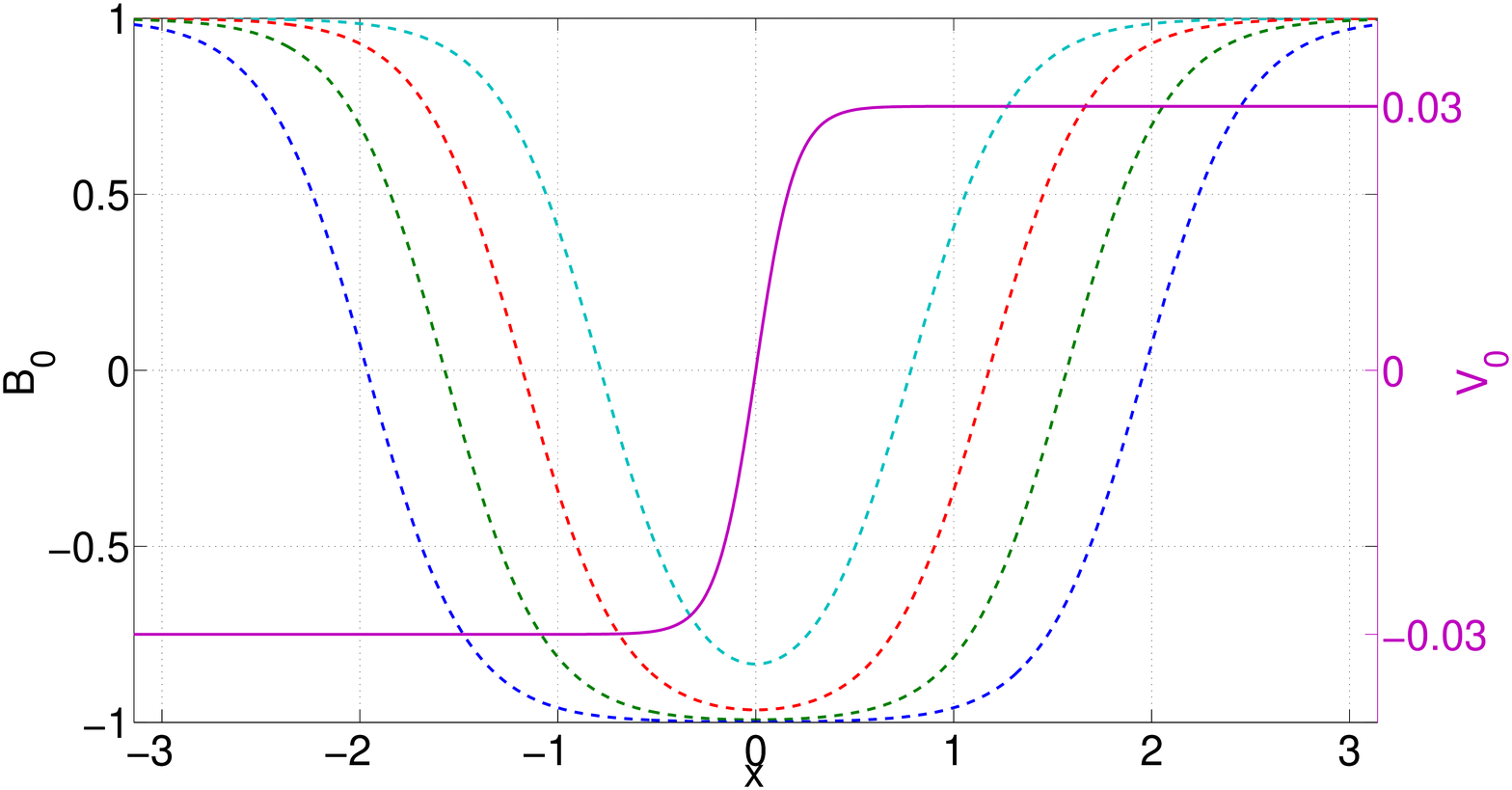}
  \caption{Equilibrium profiles of the magnetic (dased lines) and velocity fields (solid line).}
  \label{fig:profiles}
\end{figure} 
We use a two field model corresponding
to a reduced (MHD) description of the fluid
equations that provides a minimal framework
to study the impact of a radially sheared flow $\phi_0(x)$ on a double tearing mode. 
 The model consists of a set of two
coupled equations for the fluctuations of the electrostatic potential $\phi$ and magnetic flux $\psi$. 
The equilibrium magnetic field in the $z-$ direction is given by the constant $B_{0z}$ and in the poloidal by ${\bf B_0}(x)=\psi_0'(x){\bf y}$. The equilibrium current is therefore $j_0=\psi_0''(x)$. 
The time evolution of the two fields is described by
%
\hskip-2truecm
\begin{eqnarray}
  \label{RMHD00}
     \partial_t \omega + [ \phi +\phi_0, \omega +\omega_0] &=& [ \psi+\psi_0 , j+j_0 ] \nonumber \\ 
                                                                                                           & & +\nu \nabla_{\perp}^{2}\omega\;,\\
        \partial_t \psi + [ \phi+\phi_0 , \psi+\psi_0 ] &=& \eta {j} \;,
\end{eqnarray}
where  $\eta$ is the resistivity, $\nu$ the viscosity, 
$\omega=\nabla_\perp^2\phi$  the vorticity, and $j= \nabla_\perp^2\psi$
the current density fluctuation. The equations are normalized by
$\tau_A=L_\perp/v_A$ for the time, $L_\perp B_{0z}$ for $\psi$, and 
$L_\perp v_A$ for $\phi$ where  $\tau_A$ is the Alfv\'en time and 
$L_\perp$ is a magnetic shear length.

The poloidal equilibrium magnetic field $B_0(x)$ is chosen as
\begin{equation}
  \label{eq:Bequil}
  B_0 = \tanh\Big(\frac{x-x_{1r}}{a_B}\Big)-\tanh\Big(\frac{x-x_{2r}}{a_B}\Big)\;,
\end{equation}
where the parameter  $a_B=0.5$ controls the width of the profile, $x\in[-L_x/2, L_x/2]$ and ($x_{1r}$, $x_{2r}=-x_{1r}$) are the locations of the resonant surfaces where double tearing instabilities develop.  The poloidal equilibrium shear flow is given  by
\begin{equation}
  \label{eq:Vequil}
  v_0 = A_v\tanh(x/a_v)\;,
\end{equation}
where $A_v=0.03$ and $a_v=0.2$. The resulting profiles are shown in fig.~(\ref{fig:profiles}). The dashed lines represent four typical equilibrium magnetic profiles we have used, corresponding to $\delta x=x_{2r}-x_{1r}=\{\pi/2, 3\pi/4, \pi , 5\pi/4\} $ (cyan, red, green and blue respectively). The purple curve indicates $V_0$.  
Equations (1--2) are solved numerically using a finite difference scheme in the x-direction, including an Arakawa algorithm for an accurate
conservation of the Poisson brackets [.,.] and a pseudo-spectral method in the $y-$direction, including an appropriate de-aliasing
scheme\cite{Arakawa97,Muraglia09a}. The number of grid points in the x direction range from $N_x=256$ to $N_x=2048$,  with a typical spatial resolution of the order $dx=5.10^{-3}$. $L_x$ and $L_y$ are the box size in the $x$ and $y$ directions respectively. 
We find that these different profiles give rise to very different dynamics. For instance, when $\delta x=5\pi/4$,  no global reconnection occurs and complex nonlinear dynamics linked to the poloidal rotation of the islands are  observed. When $\delta x=\pi/2$, islands do not rotate and there is a global reconnection process in between the resonant surfaces.  In this paper, we focus mainly on the latter situation.

\section{\label{sec:3} Coexistence of linear instabilities}
\begin{figure}[tb]
  \centering
 \includegraphics[width=7cm,clip]{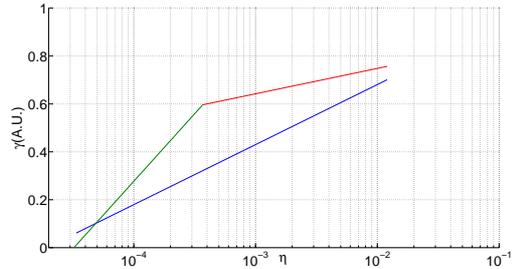}
  \caption{Scheme for $\alpha$ calculation: Without shear flow (blue line), $\gamma(\log(\eta))$ is a line and $\alpha$ the associated slope. With a shear flow, two different branches exist. For the low resistivity branch, KH is unstable (green line) and for the  high resistivity one, KH is stable (red line).}
  \label{fig:alphascheme}
\end{figure}
%
\begin{figure}[tb]
  \centering
  \includegraphics[width=7cm,clip]{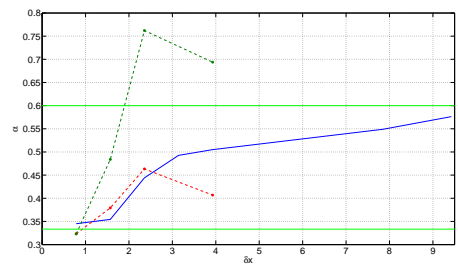}
  \caption{$\alpha$ as a function of the distance between rational surfaces $\delta x $.}
  \label{fig:alpha}
\end{figure}

It is well known that a constant magnetic field stabilizes a shear flow instability when the initial flow is a vortex sheet \cite{Biskamp00}, as far as the Alfv\'en 
velocity exceeds the amplitude of the vortex sheet. In \cite{Ofman92}, it is shown numerically that when the flow  is forced with a velocity field ${\bf v}(x) = A(1/\cosh(x/a)-1){\bf y}$, KHI coexist with the magnetic tearing  whenever $a<1$. In the latter work, the double tearing mode in the presence of a shear flow in cases where no KHI develops has also been studied, while in \cite{Bierwage07} this has been done in a case where KHIs are present.  However, the conditions at which KHI and  DTM coexist are not yet clear.

It is instructive to focus on the role of both the distance between the resonances $\delta x$ and the resistivity for a fixed amplitude of the flow $A_v$. In the absence of shear flow, the growth rate of the resistive DTM scales with the resistivity like  $\eta^\alpha$ with $\alpha=3/5$ in the limit of very ideally stable MHD conditions and $\alpha=1/3$ when the system enters an ideally marginally stable regime. Ideally Marginally stable means that the free magnetic energy of the ideal mode $\lambda_H$ is zero\cite{Biskamp00,Ara78}. In the limit of closed resonant surfaces, the latter can be linked to the tearing instability parameter $\Delta'$. $\lambda_H$ depends on the magnetic equilibrium profile and therefore on $\delta x$. 
Fig.~\ref{fig:alpha} shows the dependance of growth rate on the resistivity in the case with/without shear flow. Changing $\delta x$ and $\eta$ \cite{Janvier09},  we plot the power law index $\alpha(\delta x)$ in fig.~(\ref{fig:alpha}) . As expected, without shear flow, the power law parameter $\alpha$ ranges in between the two limit cases $1/3$ and $3/5$ (blue curve). When the distance between two rational surfaces is large enough, we converge to the standard tearing law. Conversely when $\delta x$ becomes of the order of a typical magnetic shear length $L_\perp $, we enter into a full DTM regime. A global magnetic reconnection occurs nonlinearly in such cases. 

When a weak shear flow is added, we have to discriminate between two regimes. A regime in which the KHI is unstable appears, depending on the numerical value of the resistivity. This instability is radially localized in the vicinity of the layer where the velocity shear is maximal $x\sim 0$. Typically, it has a high poloidal mode number but we observe that it modifies the growth of the tearing $m=1$ mode ($k_m=m\; 2\pi/L_y$). In fact, this not surprising because the DTM regime is linked to a strong radial coupling between the two magnetic surfaces and the presence of vortices in between them modifies the nature of the interaction, and {\it a priori} should weaken it. 
In fig.~(\ref{fig:alpha}),  the high and low resistivity cases, the KH stable (red curve) and KH unstable regimes (dark green curve) are shown. We observe that, for $\delta x \le 2$, the presence of a KH instability in between the resonances amplifies the power law index strongly such that it even exceeds  the asymptotic regime $\alpha=3/5$.
On the other hand, we observe of strong decrease of the power law index for $2\le \delta x \le 4$ in the KH stable regime, wich roughly follows the case without shear flow, but is even lower. 

The reason is that the imposed shear flow shown in fig.~(\ref{fig:profiles}) ideally leads to a poloidal rotation of the island in both directions. The global poloidal rotation of the plasma modifies the energy balance of the system and in fact reduces the growth rate of the magnetic instability. The viscosity, the resistivity, and also the nature of the mode can prohibit such an ideal scenario. In fact, the plasma starts to rotate around $\delta x \sim 2$, but when the resonances are closer together, given the weakness of the amplitude of flow $A_v$ and the strong interaction between the growing islands, the plasma poloidal rotation is locked.

\section{\label{sec:4} Structure of the m=1 DTM and generation of a mean poloidal flow }
\begin{figure}[tb]
  \includegraphics[width=7cm,height=3cm]{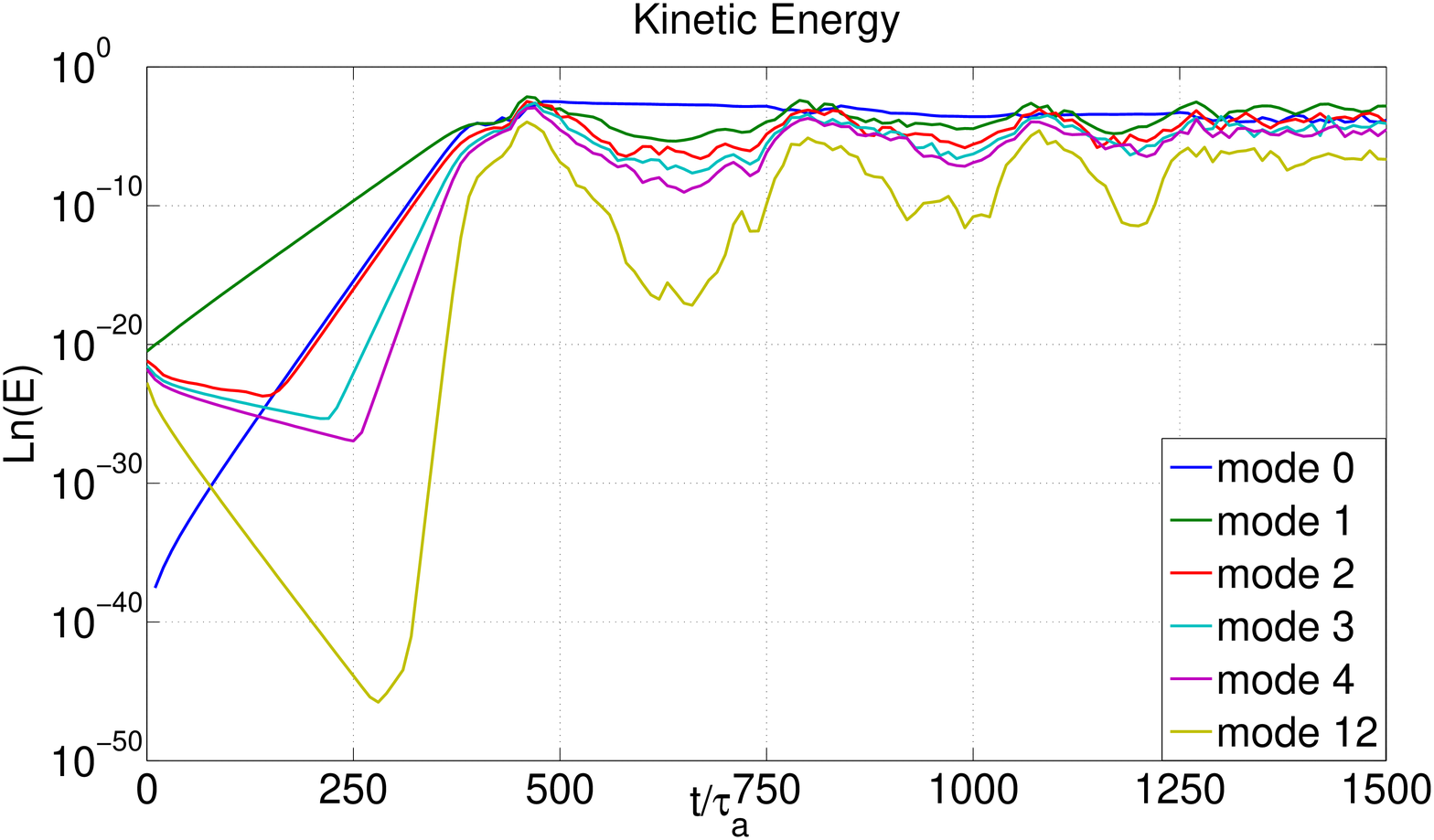}
   \includegraphics[width=7.cm,height=3cm]{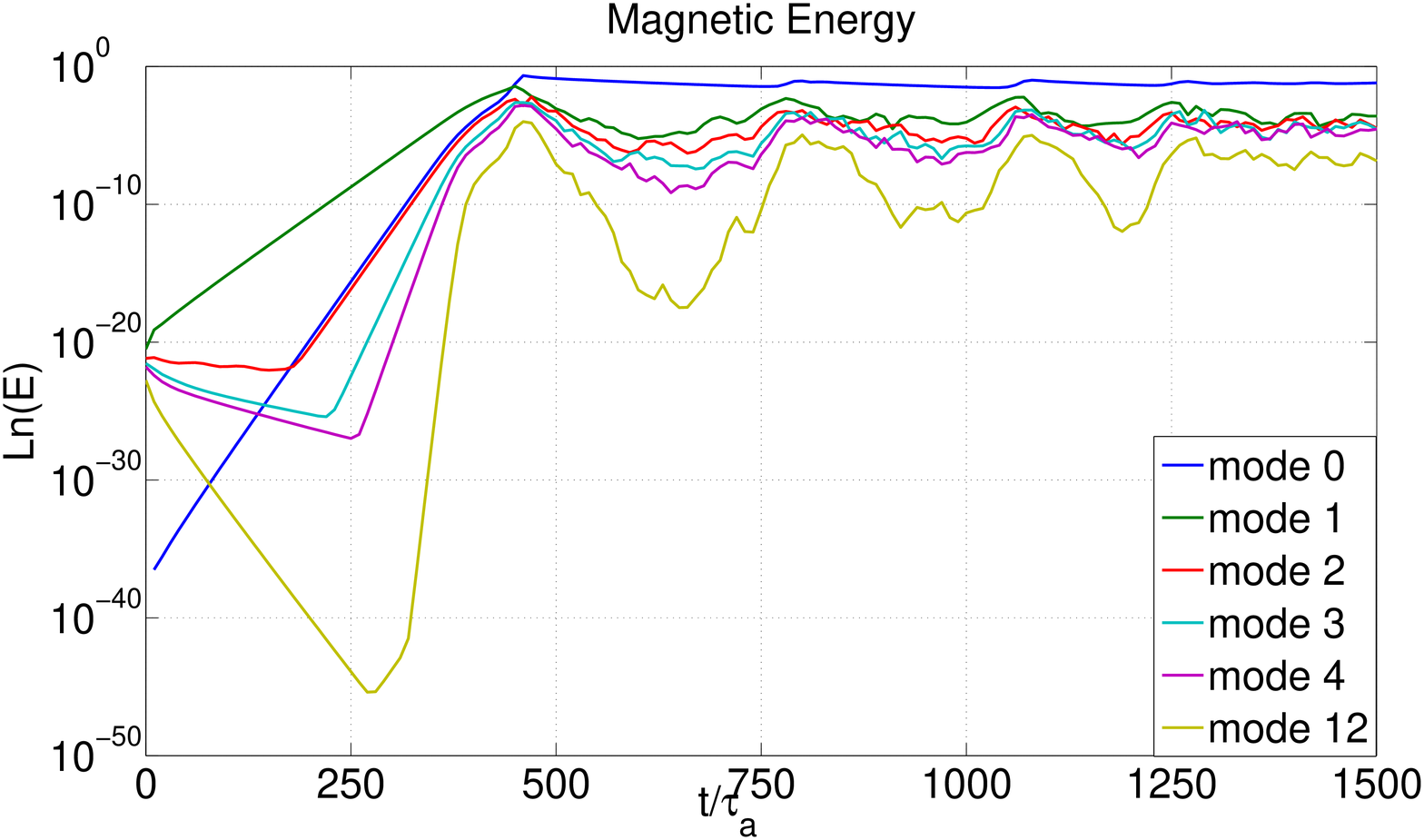}
  \caption{Energy of the poloidal modes versus time. $\delta x= \pi/2$.}
  \label{fig:EnkDTVCSf}
\end{figure}
In this section we focus on the nonlinear evolution of the m=1 DTM. Fig.~\ref{fig:EnkDTVCSf}  shows the time evolution of the kinetic and magnetic
energy of the $m=\{0,1, 2, 3, 4, 12\}$  modes for the case with $\delta x= \pi/2$, $\eta=10^{-3}$, $\mu=10^{-4}$, $L_x=2\pi$, and $L_y=2\pi$, where KH is linearly stable and only the m=1 DTM is unstable. The $m>1$ modes are stable in the linear phase ($t<160$) and are destabilized in the quasilinear phase, where the exponential growth of the modes satisfies the relation $\gamma_m = m \gamma_1$ ($t<400$) .
\begin{figure}[tb]
  \includegraphics[width=4cm,height=4cm]{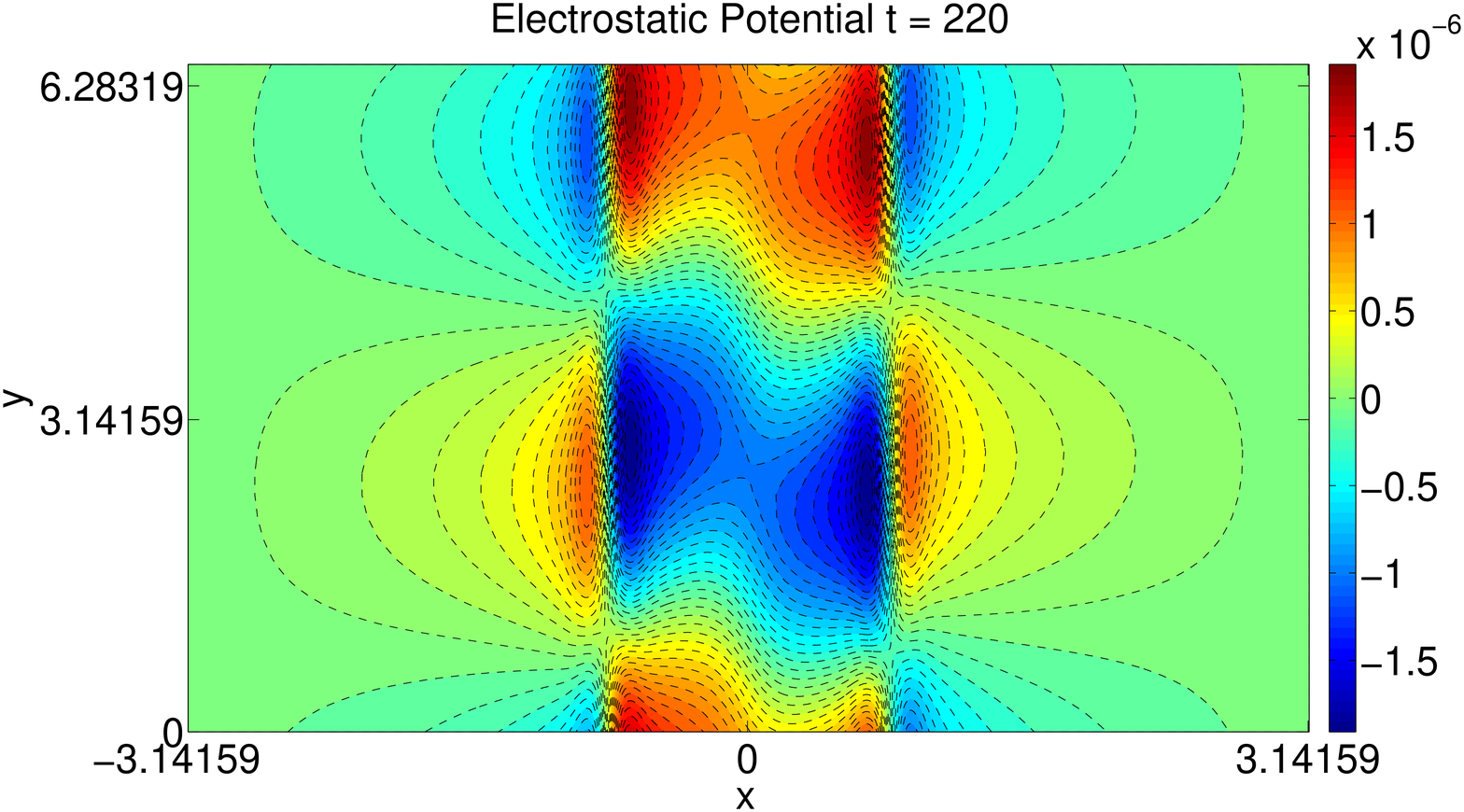}
   \includegraphics[width=4cm,height=4cm]{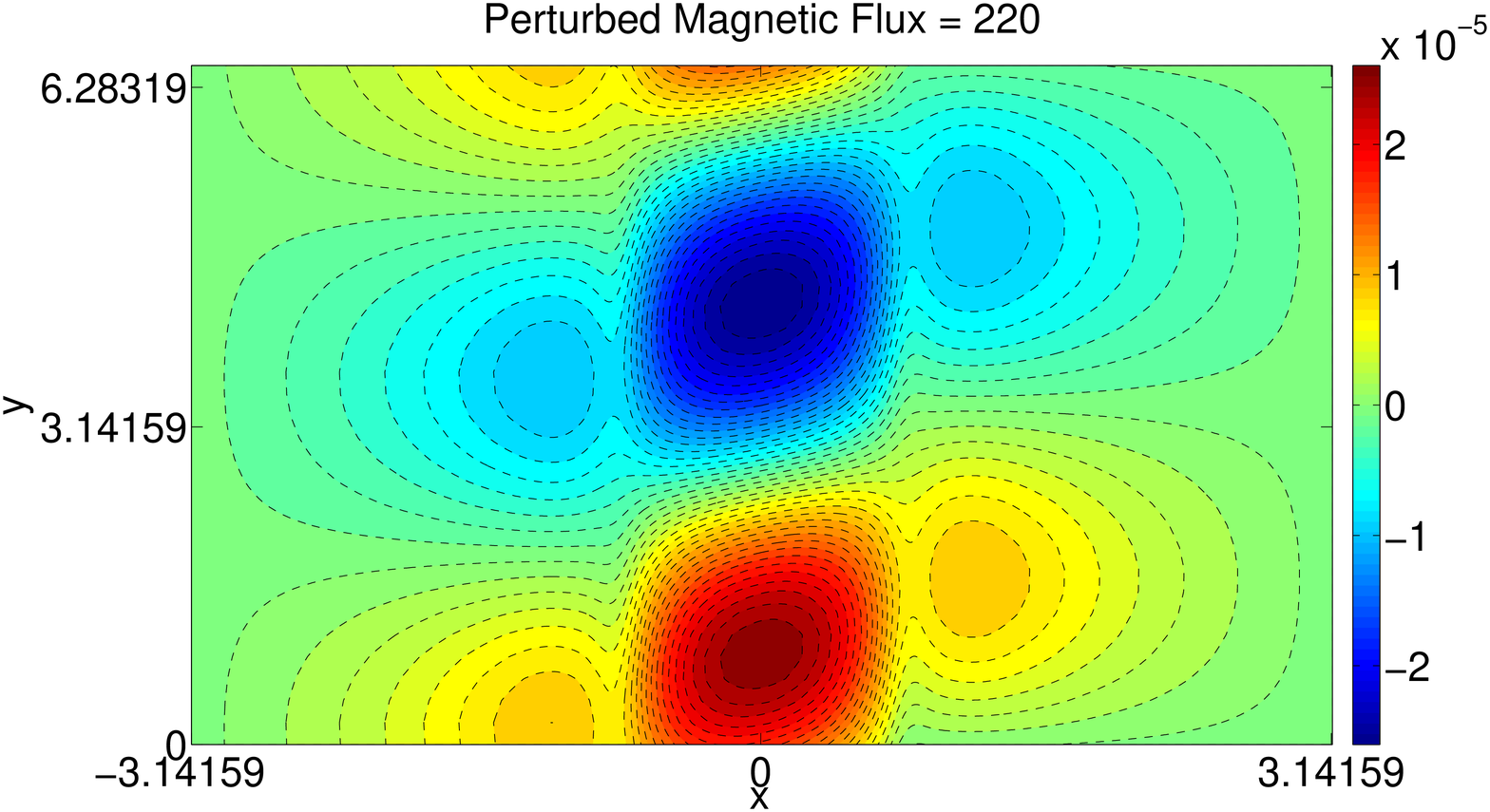}
  \caption{Snapshots of $\phi$ (Left) and $\psi$ (Right)  at $t=220\tau_A$. $\delta x= \pi/2$.}
  \label{fig:snapshots220}
\end{figure}
Contrary to the $v_0=0$ case where the $m=0$ mode is not generated in the linear and quasilinear phases, the relation $\gamma_0 = 2 \gamma_1$ is held in those phases.
We will see that the generation of the $m=0$  mode modifies the dynamics of the system leading to the global reconnection in between the resonances. In order to understand the origin of this mode, it is instructive to observe the snapshots of the electrostatic potential and the magnetic flux in the quasilinear phase. They are shown in fig.~(\ref{fig:snapshots220}).  Note that the structures are not symmetric with respect to a reflection at  $x=0$ and/or $y=\mbox{constant}$ 
as discussed in the previous section. Despite an imposed poloidal shear flow, the islands do not rotate poloidally. However,
a symmetry is clearly broken when compared with the $v_0=0$ case. Indeed, in the latter case, the $m=1$ mode satisfies a symmetry $\phi_1(x,-y,t) =-\phi_1(x,+y,t)$ where
\begin{eqnarray}
\phi_1(x,y,t)&=&\hat\phi_1(x,t) \cos(\alpha_1(x)+k_1y)
\end{eqnarray}
is the $m=1$ mode, and $\hat\phi_1$ and $\alpha_1$ are the amplitude and phase of the mode respectively. In other words, when $v_0=0$,  $\alpha_1$ is constant because of the symmetry. This is clearly false when $v_0\ne 0$ as is seen  in fig.~\ref{fig:snapshots220}. In fact, a straightforward calculation shows that whenever $\alpha_1$ is constant, the projection of the Poisson brackets $[\phi_1,\omega_1]$ on $m=0$ mode is zero. for the same reason the projection of $[\psi_1,j_1]$ will also be zero when $v_0=0$. From the breaking of poloidal parity as soon as $v_0\ne 0$, it follows that the generation of the $m=0$ mode occurs at, of course, a growth rate $2\gamma_1$ through both the Reynolds and Maxwell stresses.

\section{\label{sec:5} Impact of the shear flow on the  global reconnection process}
%
\begin{figure}[tb]
  \centering
  \includegraphics[width=4cm,height=4cm]{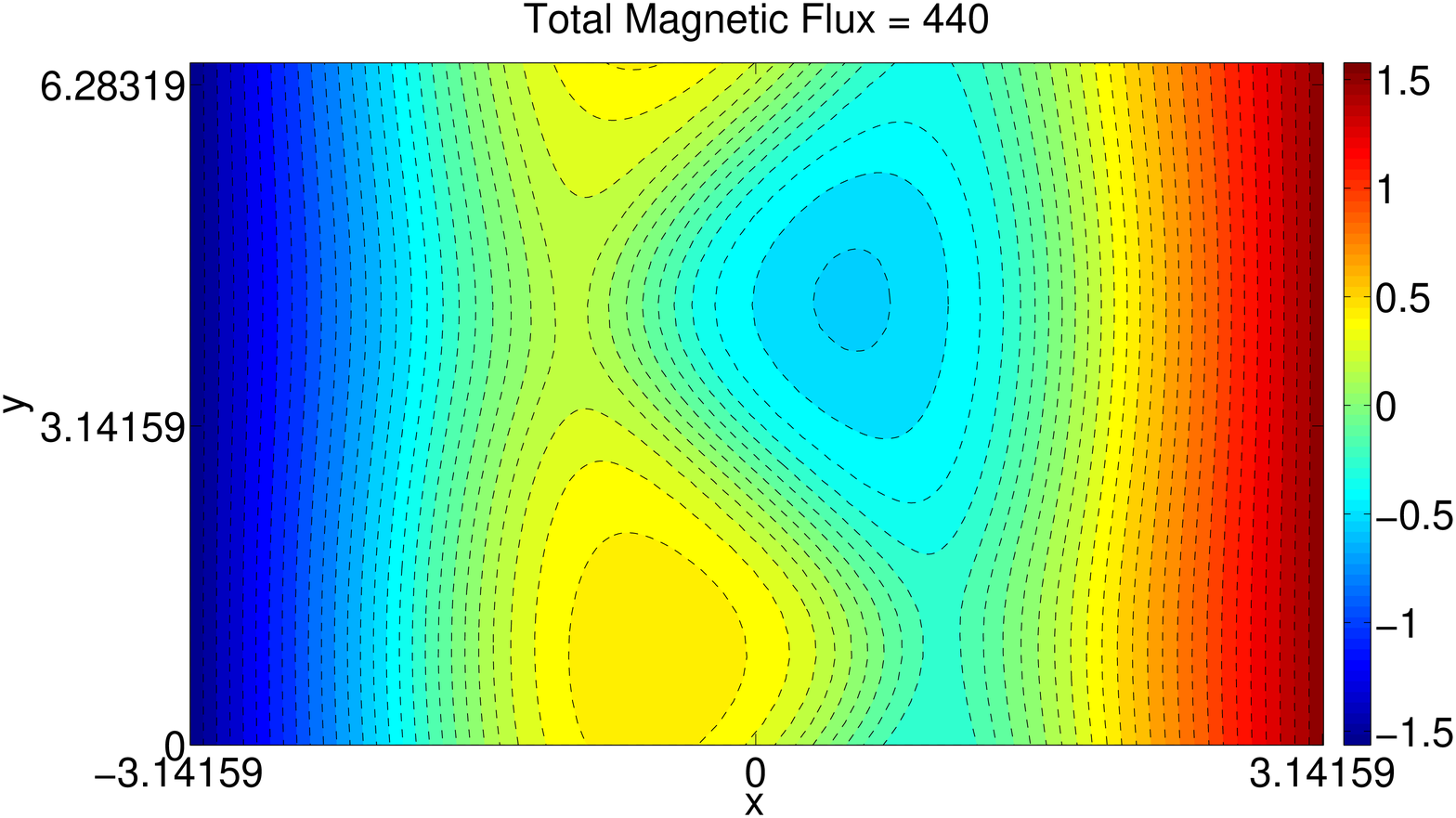} 
  \includegraphics[width=4cm,height=4cm]{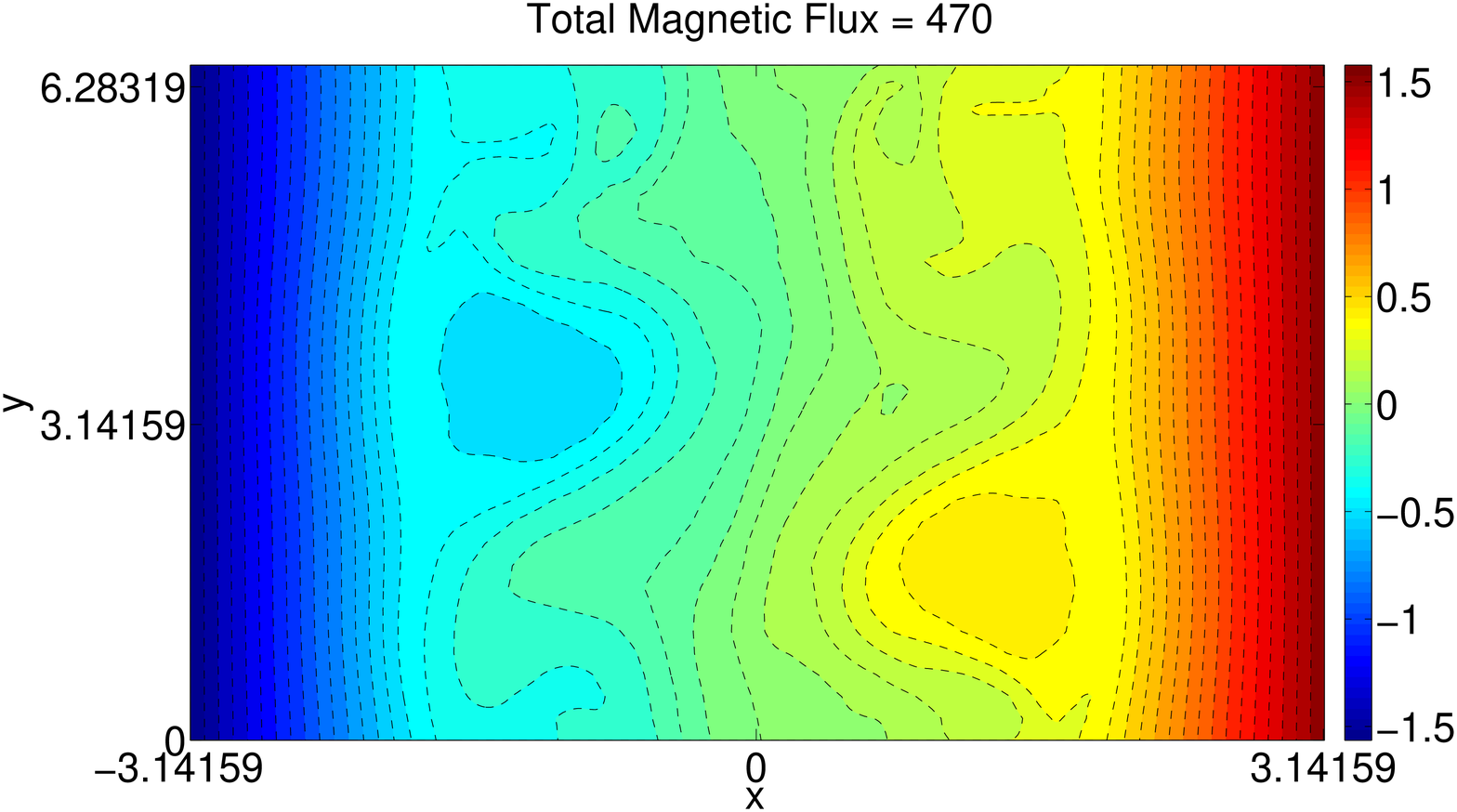} 
   \caption{Total magnetic flux just before and at the end of the global reconnection process. $\delta x= \pi/2$. (Left) $t=440\tau_A$. (Right) $t=470\tau_A$.}
    \label{fig:Total_mag}
\end{figure}

When the islands are sufficiently closed together, the global reconnection occurs regardless of a shear flow, as is seen in  fig.~\ref{fig:Total_mag}
 where  the nonlinear interaction of the two magnetic islands deforms the DTM structure ($t=440\tau_A$) and forces them to vanish progressively ($t=470\tau_A$). It is interesting to check if this process is linked to the generation of the mean velocity flow fluctuation $\tilde v_0=v_0^{\mbox{\tiny tot}}(x,t)-v_0(x)$ in the quasilinear phase of the $m=0$ mode. 
 A close examination of fig.~(\ref{fig:EnkDTVCSf})  studied together with snapshots of the magnetic flux shows a correlation. Indeed, the $\hat\psi_1$ component of the $m=1$ mode starts to move radially when the kinetic energy of the $m=0$ mode, wich is linked to  $\tilde v_0$,  becomes more important than that of the $m=1$ mode (step 1), in the $\delta x=\pi/2$ case at  $t\sim395$. 
 The global reconnection process can then develop and occur in the time interval $t \in [440,480]$. In fact, when the magnetic energy of the $m=0$ mode crosses  that of the $m=1$ mode at  $t\sim448$ (step 2), the remaining islands are no longer topologically linked to their initial resonant surface (see right snapshot of fig.~(\ref{fig:Total_mag})).   At $t=478$, the global reconnection process is completed and the islands disappear (step 3).
\begin{figure}[tb]
  \centering
  \includegraphics[width=4cm,height=4cm]{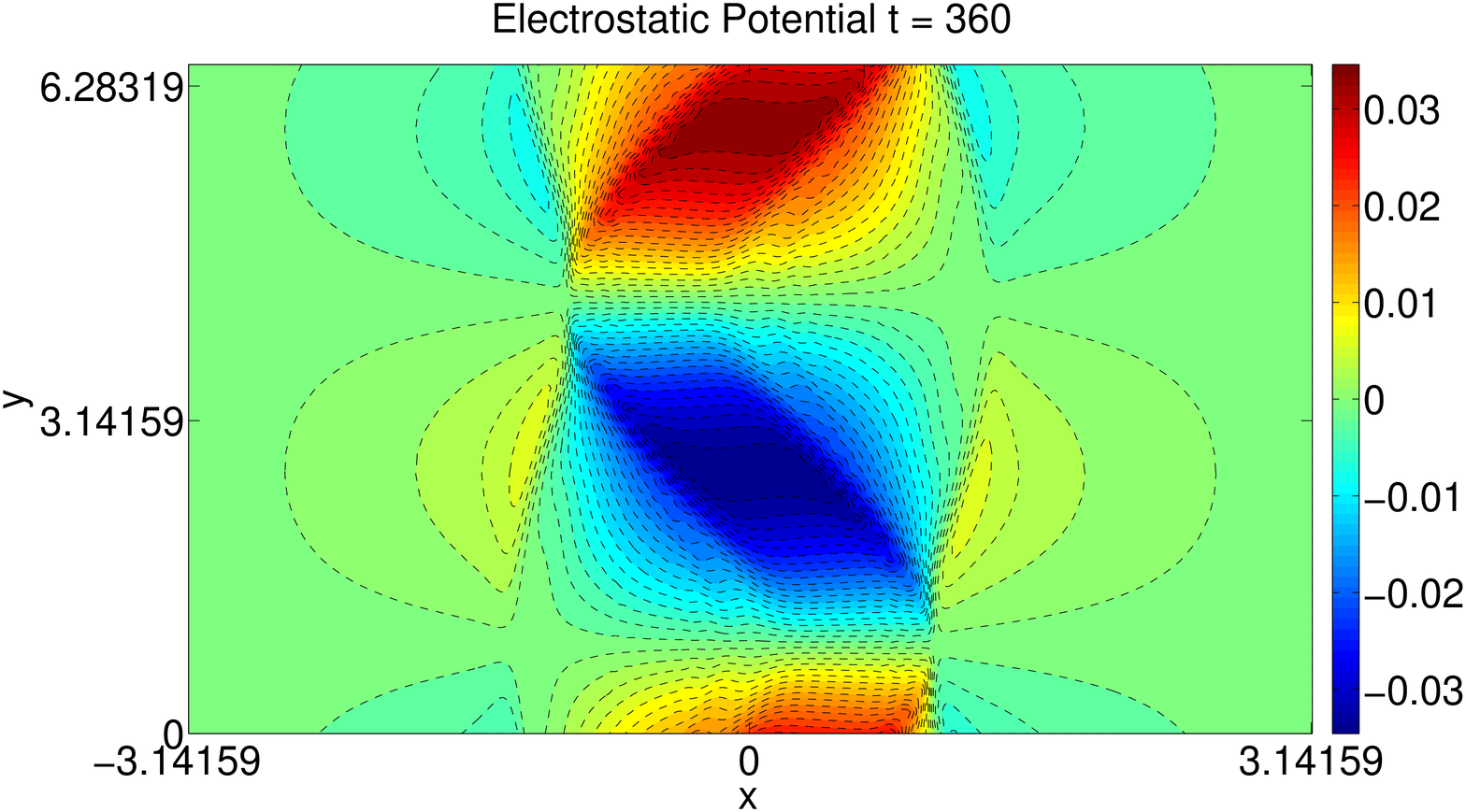} 
 \includegraphics[width=4cm,height=4cm]{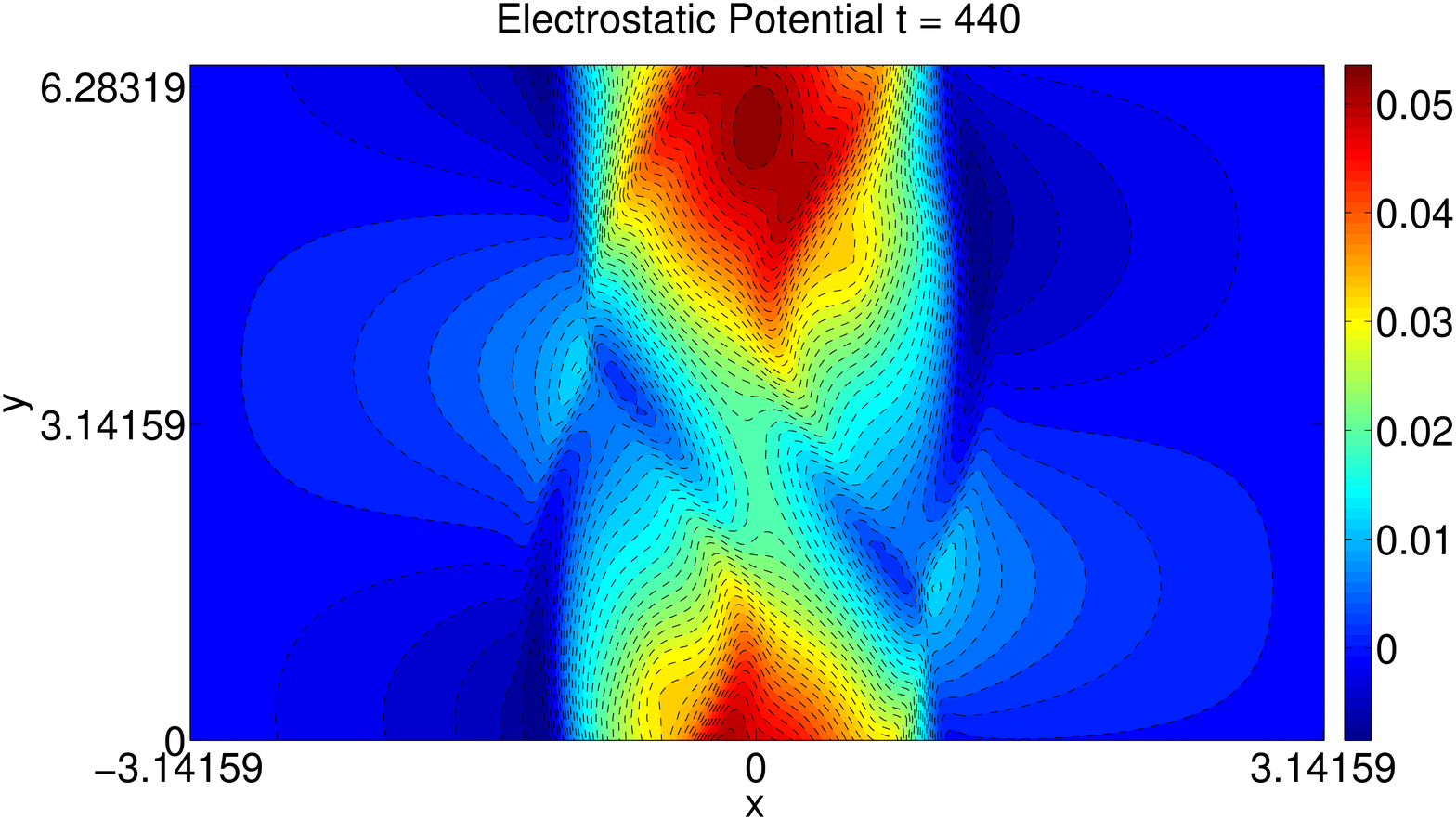} 
  \caption{$\delta x= \pi/2$. Perturbed electrostatic potential just before the global reconnection. (Left) No shear flow. (Right) With the shear flow}
    \label{fig:PerturbedEP}
\end{figure}
Inspection of the maps of the electrostatic potential just before the global reconnection (see fig.~(\ref{fig:PerturbedEP})) shows that  the presence of the 
 $m=0$ mode, even if it appears to be much more complex,  does not substantially modify the plasma flow compared with the case when $v_0=0$, where typical butterfly-like structures are easily identified. To clarify the role of the generated poloidal mean flow  $v_0^{\mbox{\tiny tot}}(x,t)$, it is interesting to focus on the time evolution of its structure.
\begin{figure}[tb]
  \centering
   \includegraphics[width=4cm,height=4cm,clip]{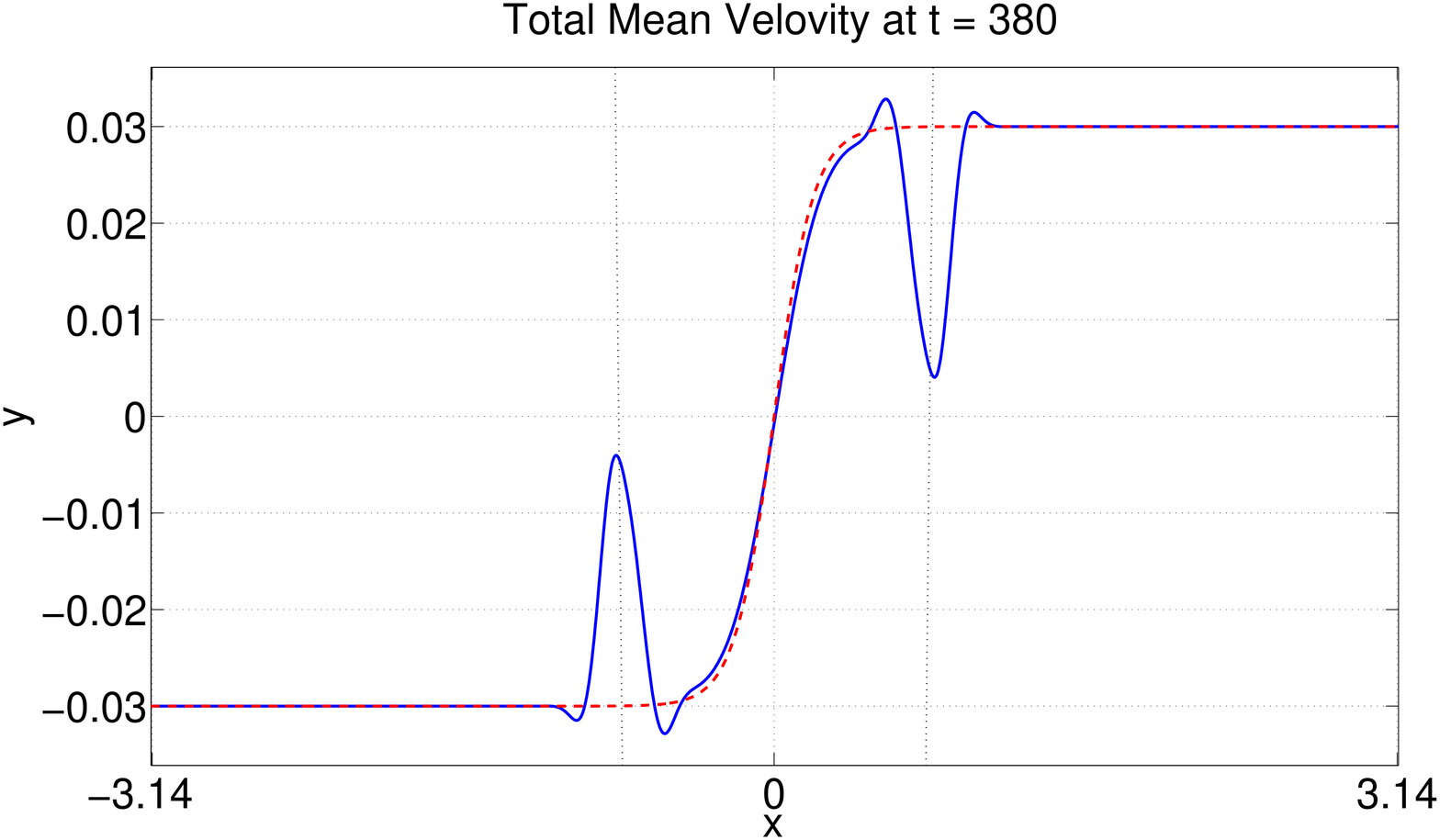}  
   \includegraphics[width=4cm,height=4cm,clip]{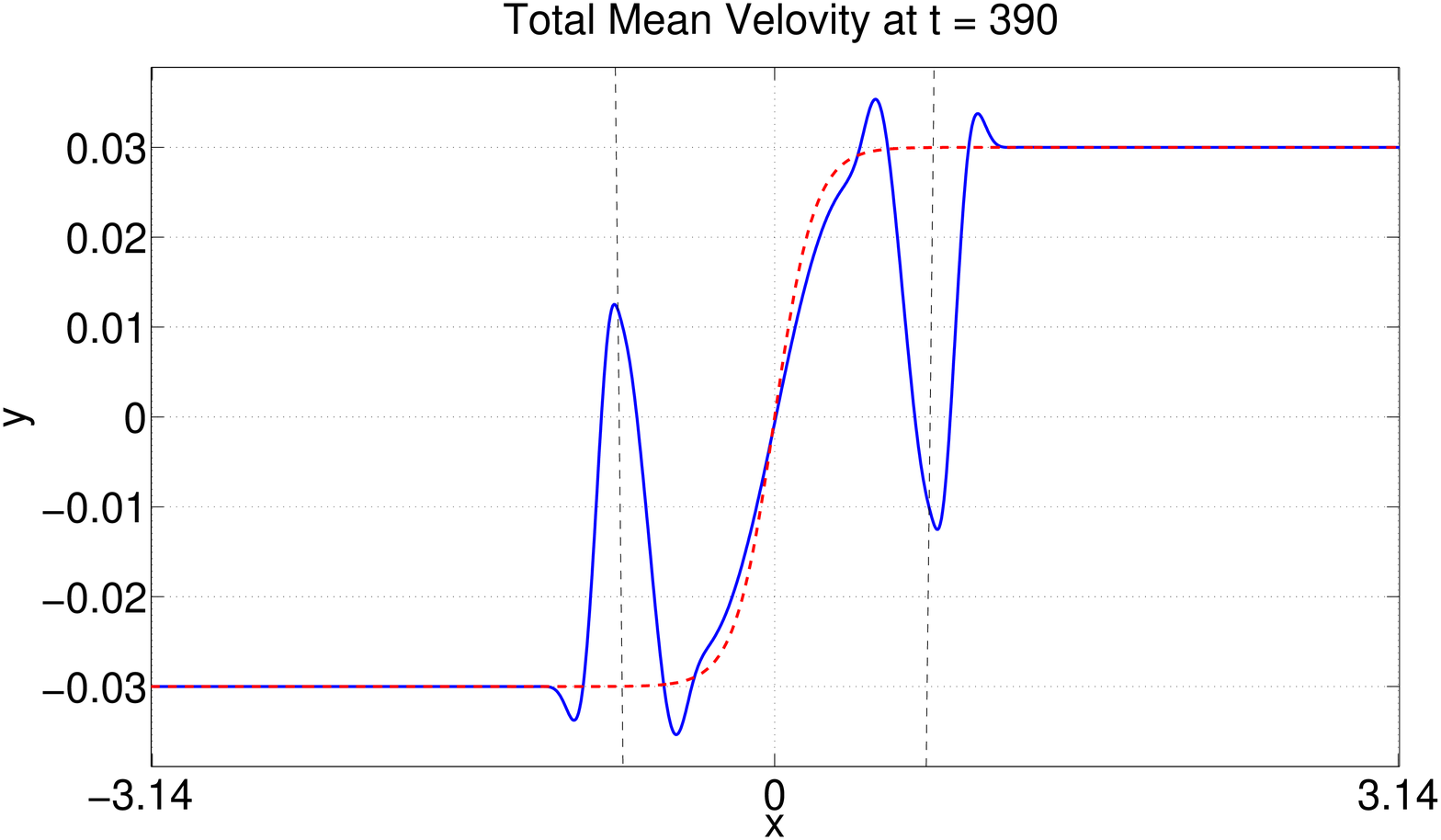}  
   \includegraphics[width=4cm,height=4cm,clip]{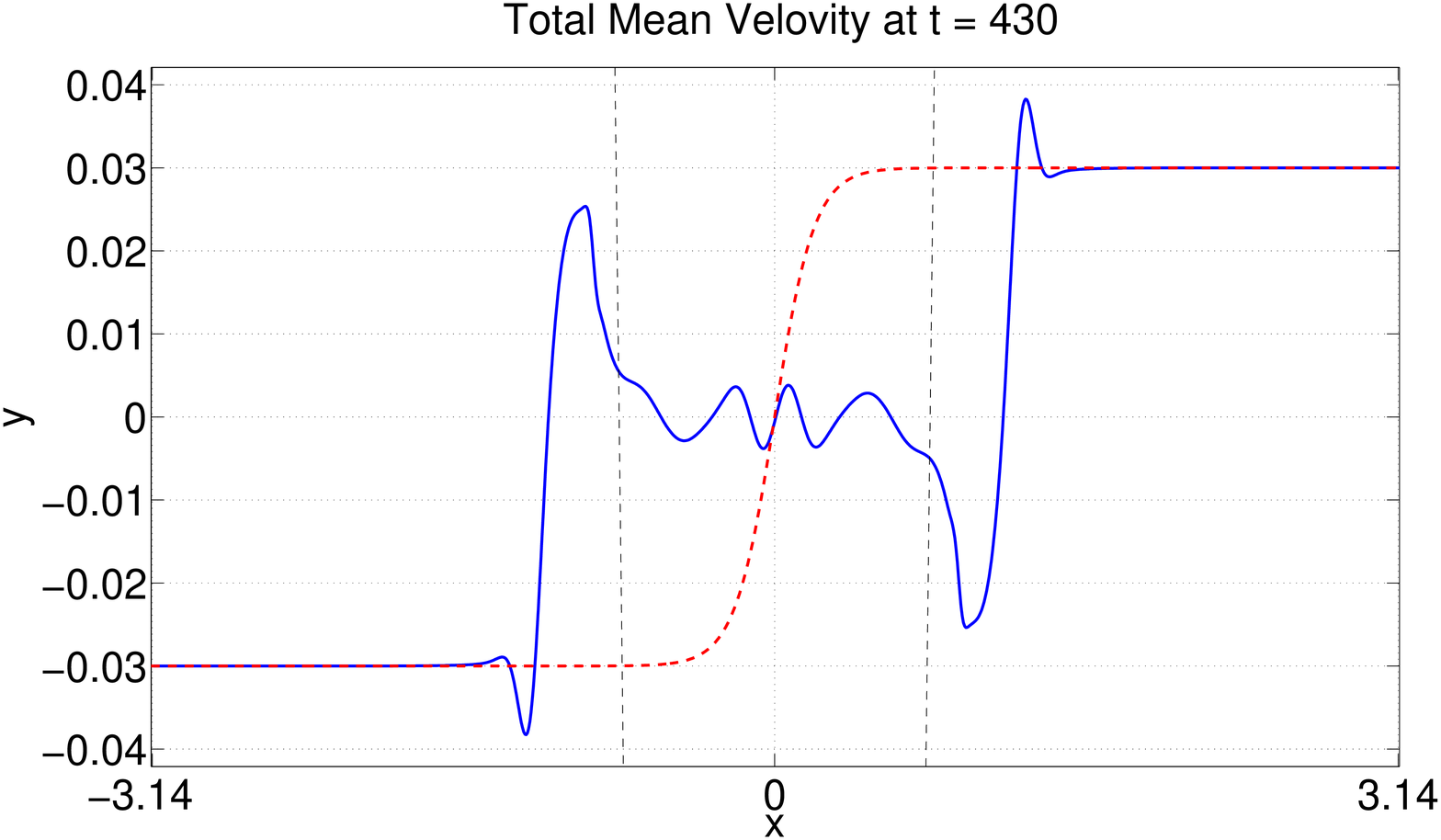} 
   \includegraphics[width=4cm,height=4cm,clip]{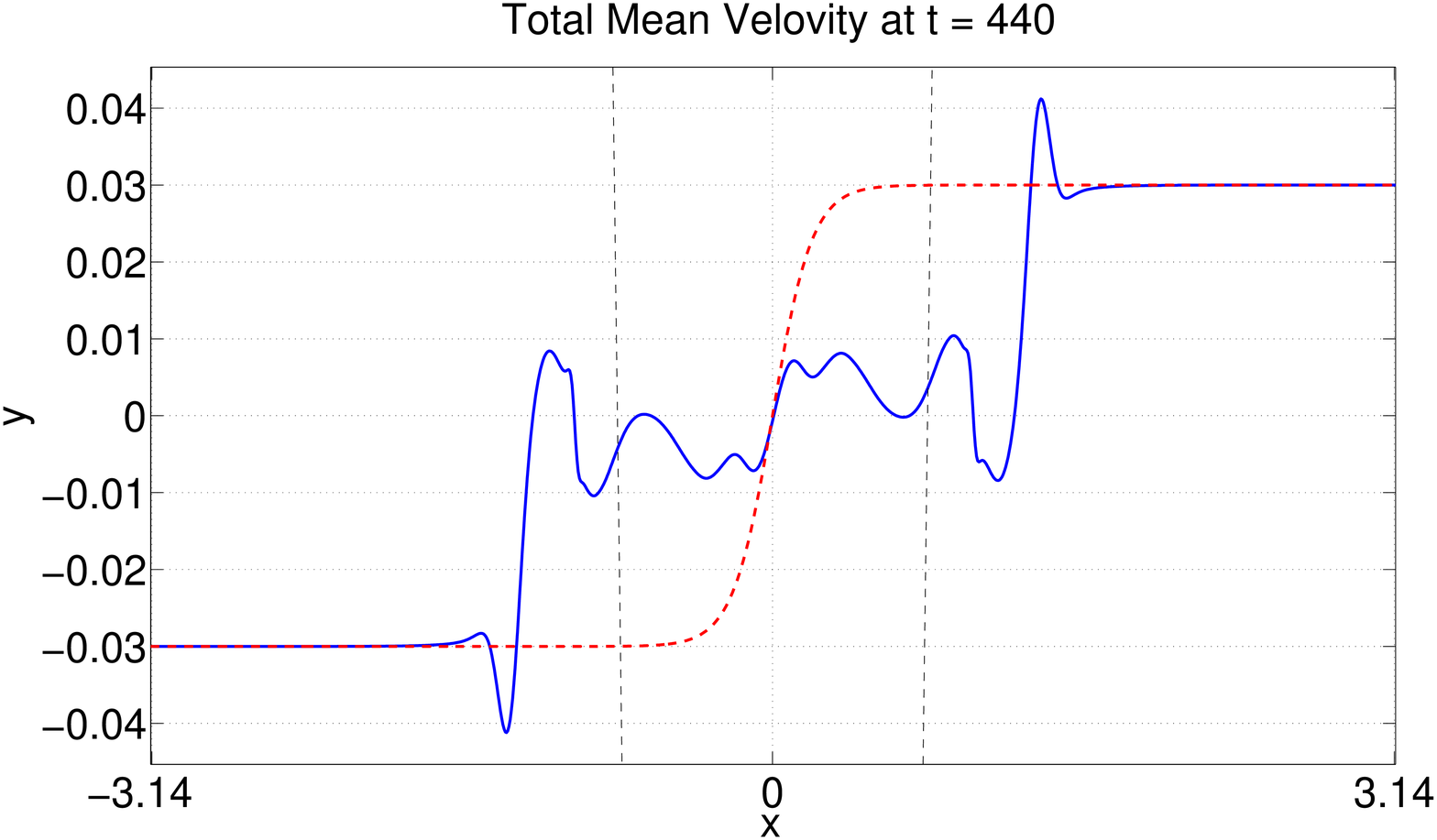}  
  \caption{$\delta x= \pi/2$. Structure of $v_0^{\mbox{\tiny tot}}(x,t)$. (Top-Left) $t=380 \tau_A$. (Top-Right) $t=390\tau_A$. (Bottom-Left) $t=430\tau_A$. (Bottom-Right) $t=440\tau_A$}
     \label{fig:v0versustime}
\end{figure}
\begin{figure}[tb]
 Ê\centering
 Ê \includegraphics[width=3.85cm,height=4cm,clip]{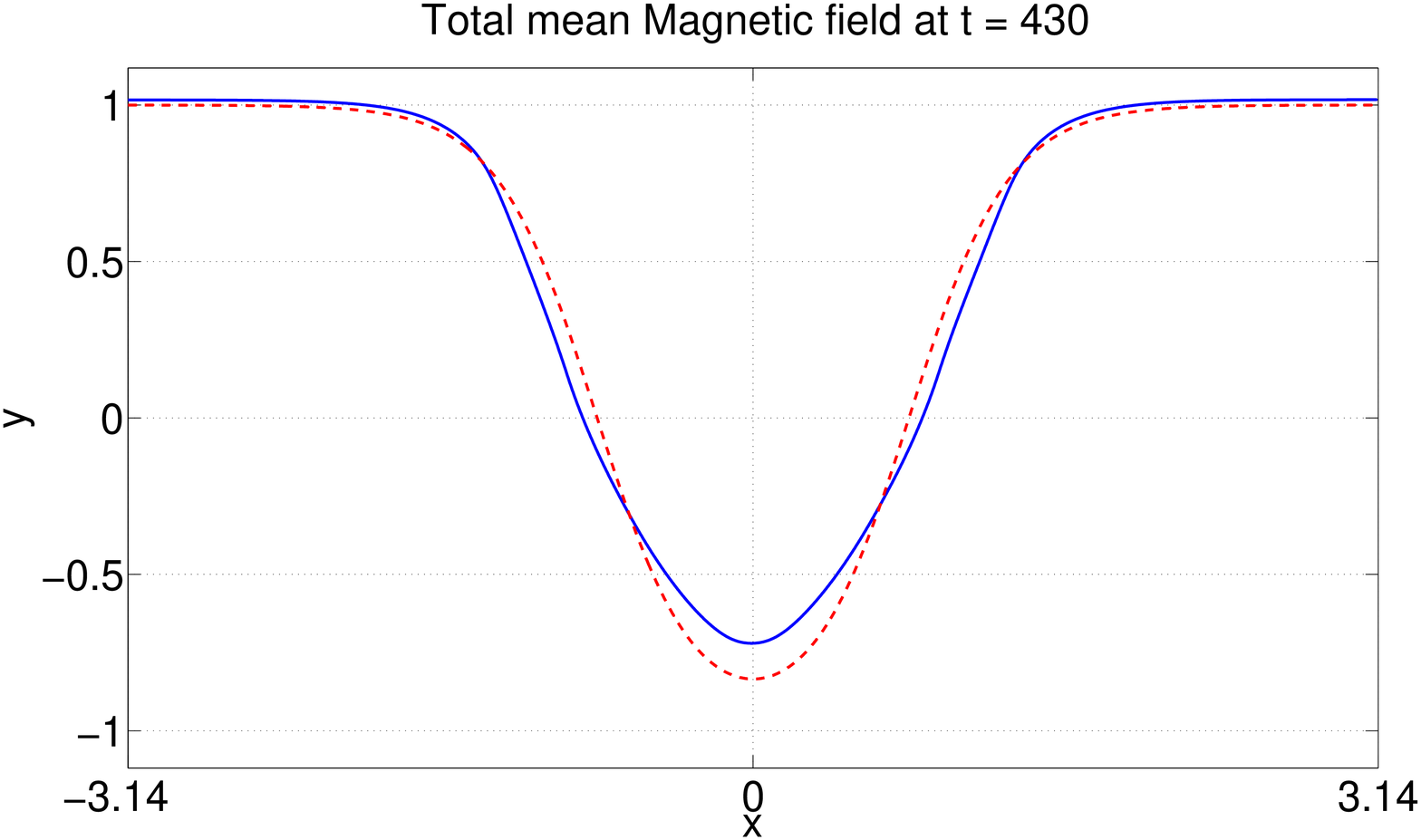} Ê
 Ê \includegraphics[width=3.85cm,height=4cm,clip]{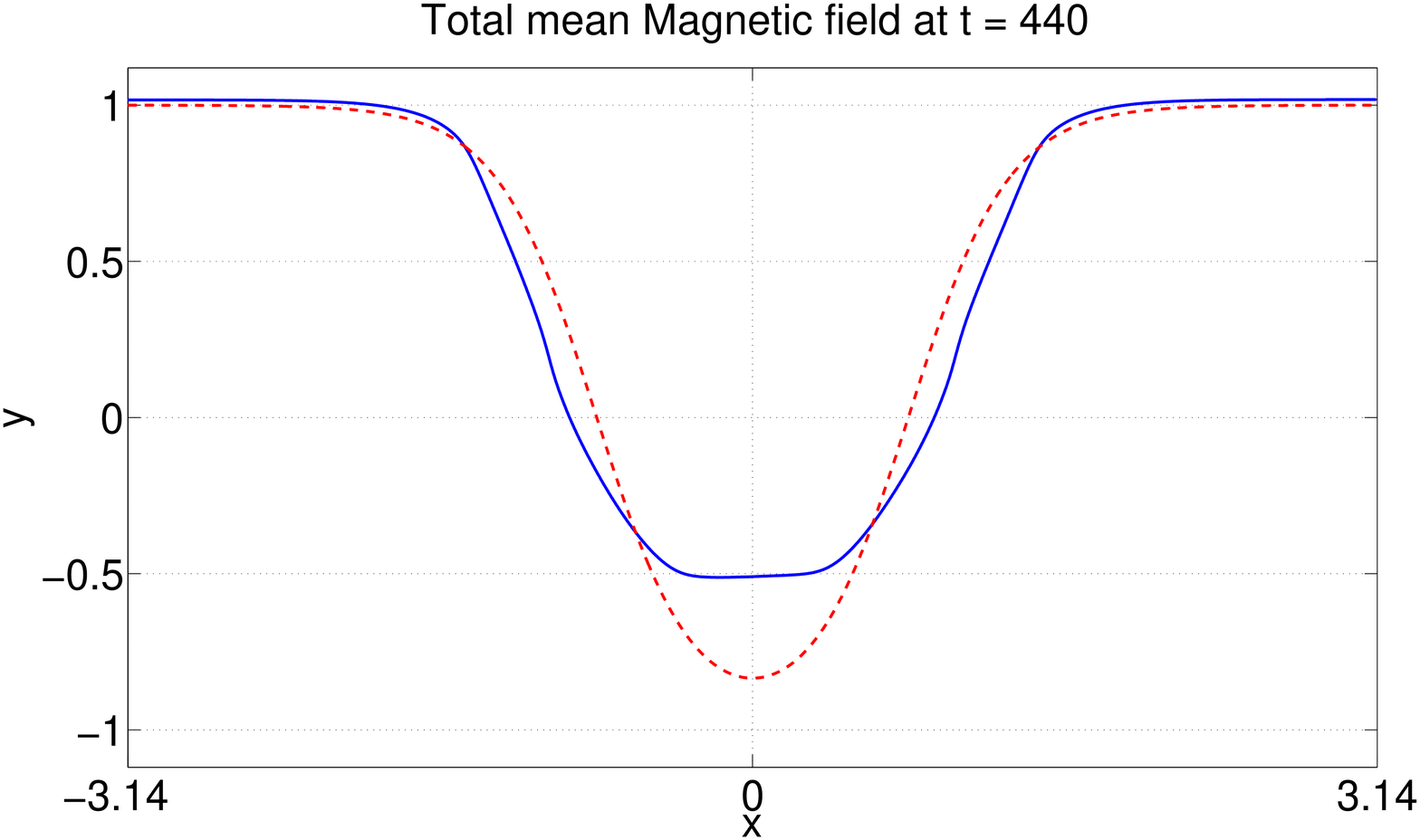} Ê
 Ê \includegraphics[width=3.85cm,height=4cm,clip]{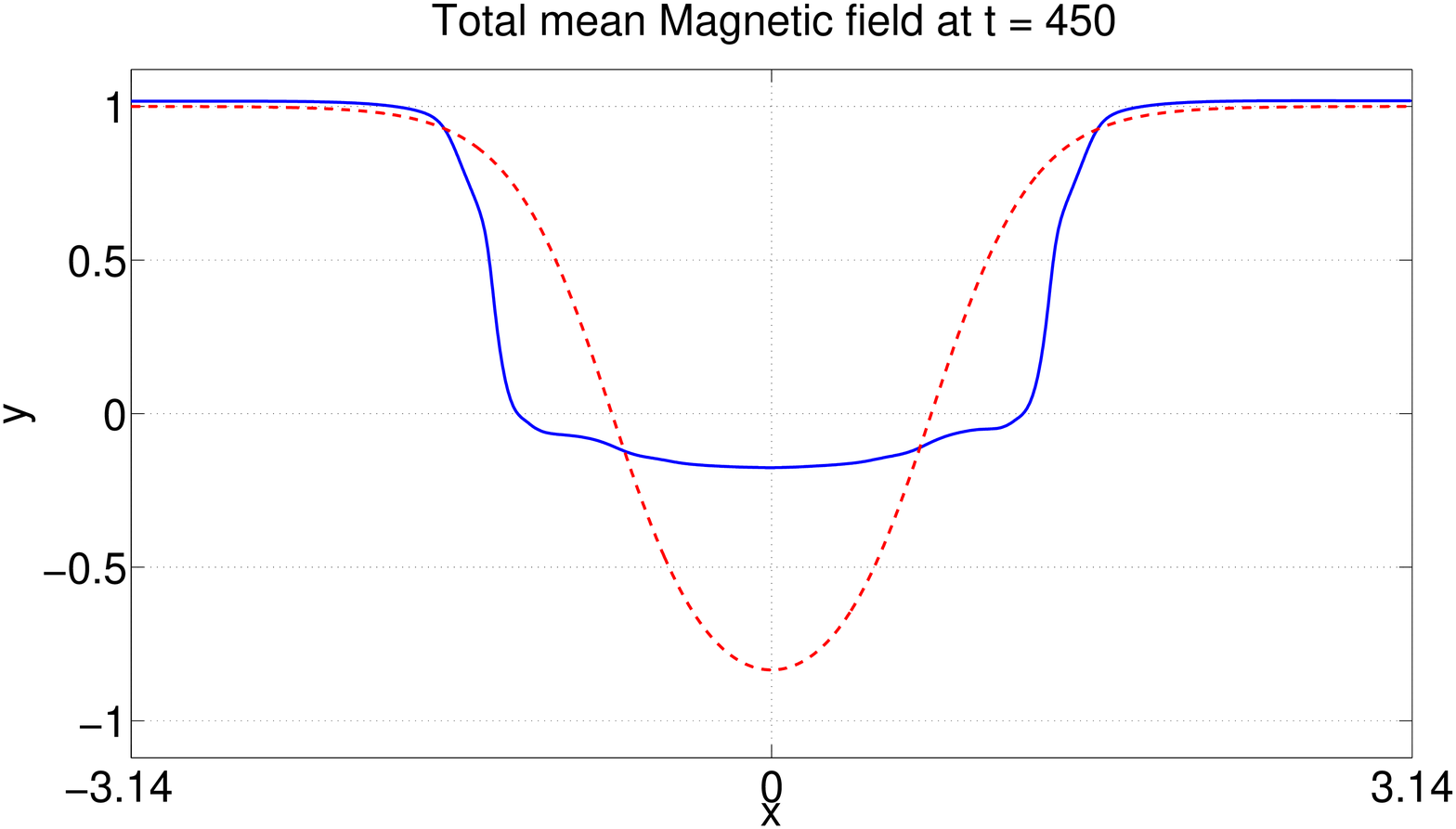} 
 Ê \includegraphics[width=3.85cm,height=4cm,clip]{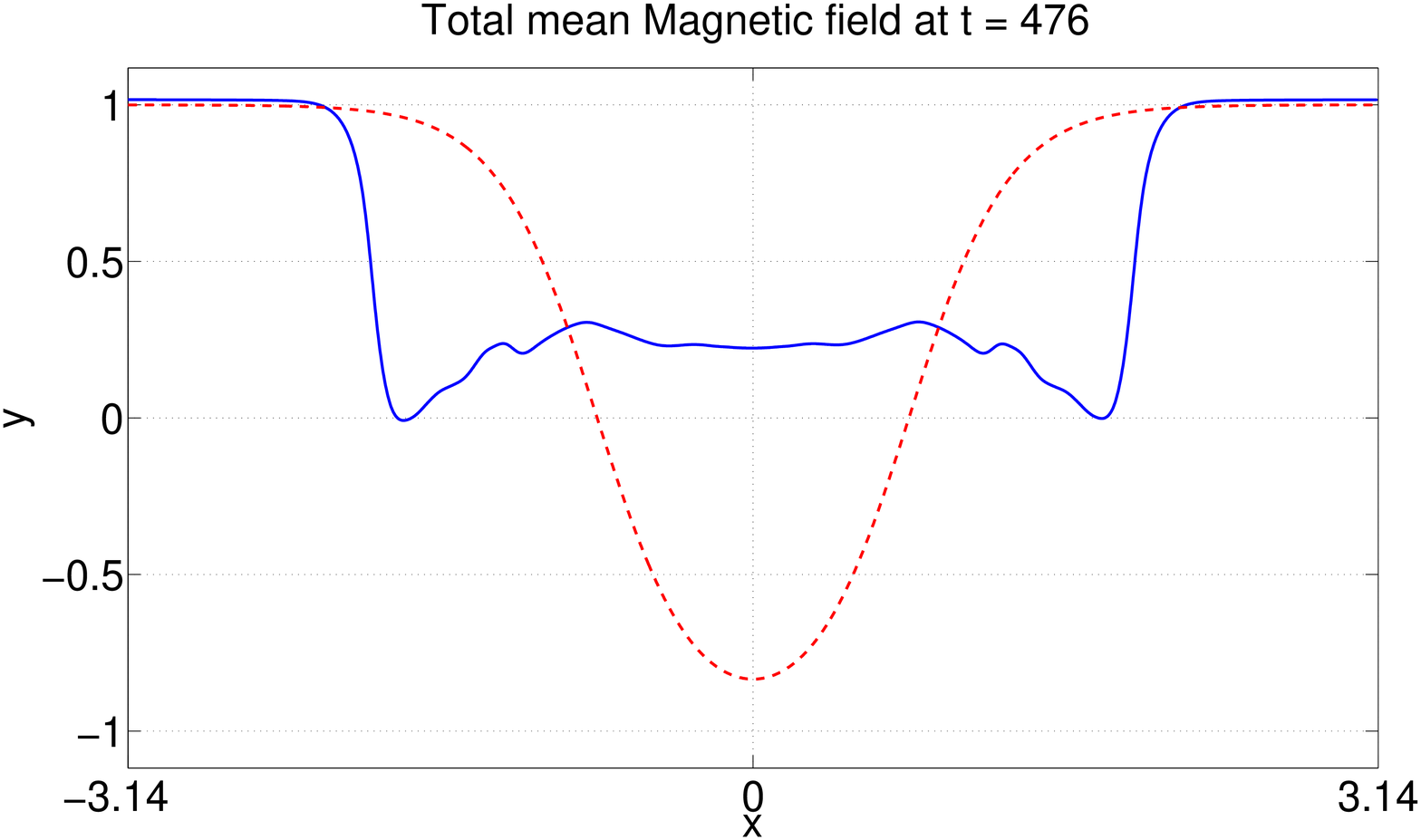} Ê
 Ê 
 Ê\caption{ $\delta x= \pi/2$. Structure of $B_0^{\mbox{\tiny tot}}(x,t)$. (Top-Left) $t=430 \tau_A$. (Top-Right) $t=440\tau_A$. (Bottom-Left) $t=450\tau_A$. (Bottom-Right) $t=476\tau_A$}
  Ê \label{fig:B0versustime}
\end{figure}
Fig.~(\ref{fig:v0versustime}) shows $v_0^{\mbox{\tiny tot}}(x,t)$ at different times.  The  red lines represent $v_0=v_0^{\mbox{\tiny tot}}(x,0)$ and the horizontal lines show the position of the resonances.  Initially, we observe that $\tilde v_0$ grows close to the resonances. Then, (step 1)   occurs roughly when $v_0^{\mbox{\tiny tot}}(x_{ir},t)$ has crossed zero (about $t=385$) or some Alfv\'en time after this has happened  (about $t=395$). Finally, the graphs $t=430\tau_A$
and $t=440\tau_A$ show that during the global reconnection, which starts at (step 2), the mean velocity oscillates around zero in between the resonances, approaching, in some sense,  the case without imposed shear flow. These observations show, first, that the nonlinear destabilization of the DTM  occurs once the generated $\tilde v_0(x,t)$  compensates $v_0$ in the vicinity of the resonant surface, and second, that the global reconnection process starts and occurs once the mean velocity profile in between the islands oscillates around zero.

It is instructive to follow the time evolution of the mean magnetic field $B_0^{\tiny\mbox{tot}}(x,t)$. Fig.~(\ref{fig:B0versustime}) shows $B_0^{\mbox{\tiny tot}}(x,t)$ at different times during the global reconnection process, {\it i.e.} in between (step 2) and (step 3). First, we can see that until (step 2), the profile and also the positions of the resonant surfaces ($B_0^{\mbox{\tiny tot}}(x,t)=0$) are roughly unchanged. This is followed by a flattening of the profile in between the resonant surfaces where the mean current decreases, while in the vicinity of the resistive layers, the profile becomes narrow signifying the generation of a strong mean current
(see snapshot at $t=440\tau_A$ and $t=450\tau_A$, respectively). This effect is amplified until (step 3) occurs. It corresponds to the time when 
 $B_0^{\tiny\mbox{tot}}(x,t)$ does not have a zero point at any location (see $t=476\tau_A$ snapshot). In other words, there is no resonant surface anymore.

\begin{figure}[tb]
  \centering
  \includegraphics[width=7cm,clip]{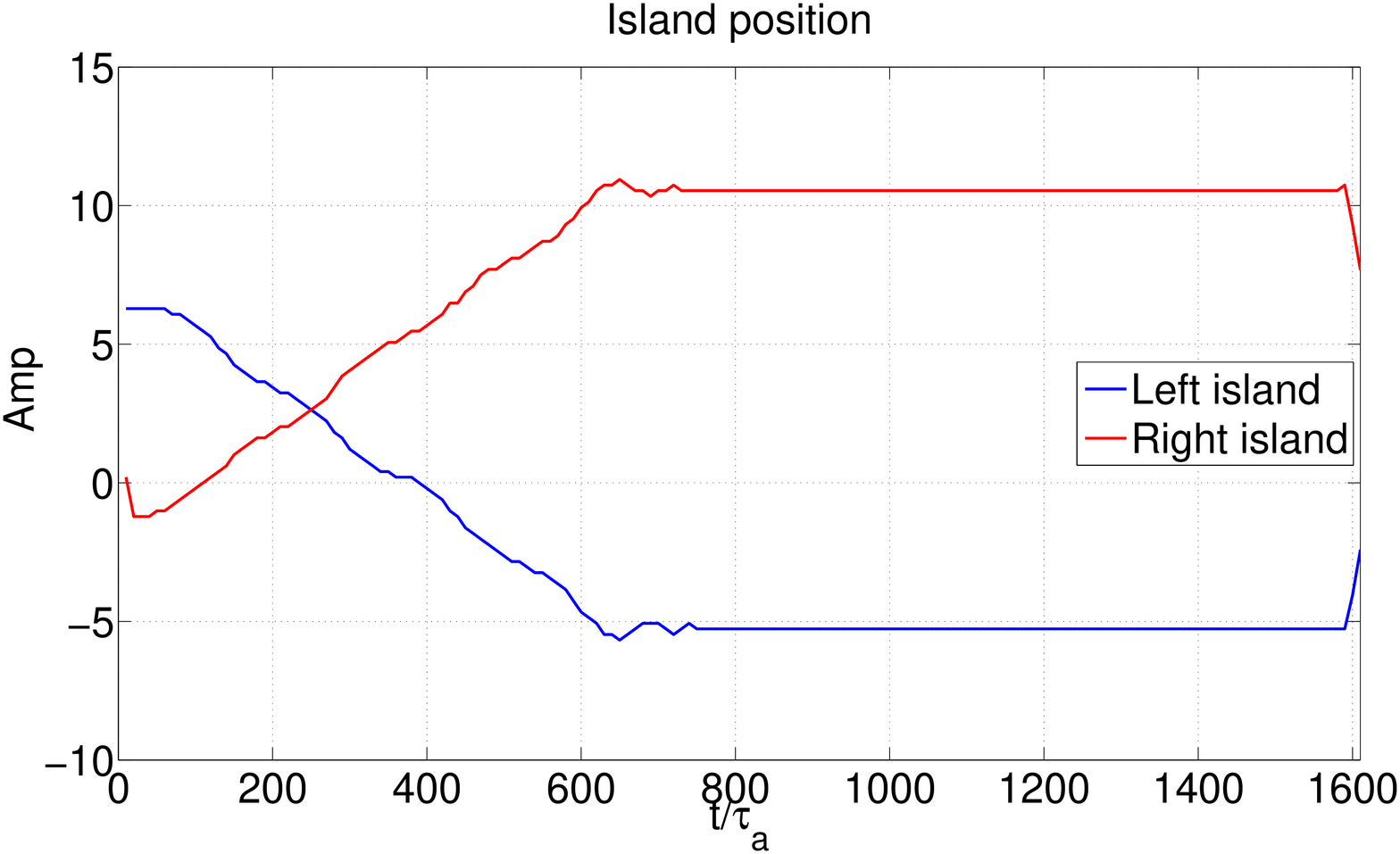} 
  \caption{ $\delta x=3 \pi/4$. Time evolution of the poloidal positions of the center of the two islands}
    \label{fig:islandposition}
\end{figure}

Let us consider also the case $\delta x= 3\pi/4$, keeping all the other parameters identical to those used in the case  $\delta x= \pi/2$. The KH stable branch still exists and shows the similar behaviour as the case with  $\delta x= \pi/2$. We find that the same behavior is observed.  However, the situation is more complex because island rotation is initially involved in the dynamics. In fact the (step 1) event occurs at $t=660$, when also the magnetic islands stop to rotate poloidally and the (step 2) event occurs at $t=1600$, after a long period where the islands do not move poloidally (see fig.(~\ref{fig:islandposition})). Moreover, in this latter case, the global reconnection corresponds to a structure-driven nonlinear instability of DTM characterized by an abrupt growth after a long-time-scale evolution\cite{Ishii02}, such that, at $t=1610$, the islands reach the wall. 
%
In both cases, however, the presence of the shear flow delays the time at which global reconnection occurs. The delay is of about $15\%$ for $\delta x= \pi/2$ and $30\%$ for $\delta x= 3\pi/4$. A much more systematic study should be performed, but this preliminary work shows that control of the global reconnection process through control of the shear flow in between the islands might be possible.
%

%
%


\section{\label{sec:8} Conclusion}

In this paper, we have presented a study of the role of a shear flow  in between magnetic islands on  the global magnetic reconnection processes.
We first have identified the different linear regimes which exist when a shear flow is present in between two surfaces where DTM develop. We have seen that according to the distance between the resonant surfaces and the resistivity of the plasma different regimes exist. When the distance is larger than the typical magnetic shear length,  no global reconnection occurs and the islands rotate\cite{Voslion08}. In this paper, we did not focus on those regimes. When it is of the order or smaller than the typical shear length of the system, an island can be locked or rotate linearly according to the value of $\delta x$. Moreover,  the system can be KH stable or unstable according to the resistivity and, of course, to the amplitude of the shear flow. We have shown that, independent of the regime (rotating island regime or not), a breaking of symmetry linked to the presence of the shear flow leads to the generation of a mean poloidal flow at early times. We have shown that flow develops first in the vicinity of the resonant surfaces and  is at the origin of the radial displacement and destabilization of the DTM structure. We have found that the presence of the shear flow delays the global reconnection process, suggesting that its control might be usefull for controlling the DTM. Further investigations are, however, necessary to precisely evaluate the impact of the shear flow and its amplitude, and also the impact of the KH instability in the low resistivity regimes, on the global reconnection processes of the double tearing mode.

{\it Acknowledgments}:
The authors wish to thank M. Muraglia and G. Fuhr for fruitful discussions. T. V. acknowledges the
College Doctoral Franco-Japonais for its fellowship. This work is partly supported by LIA 336CNRS and by NIFS/NINS under the project of formation of International Network for Scientific Collaborations, and by Grant-in-Aid for Scientific Research (21224014).

\end{document}